\documentclass{article}

\usepackage{spconf, amsmath,amsfonts,amssymb}
\usepackage{pifont}

\usepackage{multirow}
\usepackage{array}
\usepackage{verbatim}
\usepackage{booktabs}
\usepackage{tabularx}
\usepackage{longtable}
\usepackage{colortbl}
\usepackage{xcolor}

\usepackage[table]{xcolor}
\definecolor{lightgreen}{RGB}{220,255,220}
\definecolor{medgreen}{RGB}{170,240,170}
\definecolor{darkgreen}{RGB}{120,220,120}

\usepackage{graphicx}
\usepackage[font=normalsize,labelfont=bf,textfont=sf]{caption}
\usepackage{subcaption}

\usepackage{algorithm}
\usepackage{algpseudocode}

\usepackage{pgfplots}
\pgfplotsset{compat=1.16,width=10cm}
\usepgfplotslibrary{statistics,groupplots}
\usetikzlibrary{patterns, shapes, arrows, positioning}

\usepackage{adjustbox}
\usepackage{placeins}
\usepackage{url}
\usepackage{soul}  
\usepackage{listings}
\usepackage{appendix}

\usepackage[hidelinks]{hyperref}

\name{
Nitin Choudhury$^{*}$,
Bikrant Bikram Pratap Maurya$^{*}$,
Orchid Chetia Phukan$^{*}$\thanks{*Authors contributed equally as First Authors},
Arun Balaji Buduru
}

\address{
IIIT-Delhi, India \\
}

\begin{document}


\title{FOCA: Multimodal Malware Classification via Hyperbolic Cross-Attention}

\name{Nitin Choudhury$^{*}$,  Bikrant Bikram Pratap Maurya$^{*}$\thanks{$^{*}$Authors contributed equally as first author}, Orchid Chetia Phukan$^{\#}$\thanks{$^{\#}$Core Ideation and Supervision}\thanks{Corresponding Author: \href{mailto:orchidp@iiitd.ac.in}{orchidp@iiitd.ac.in}}, Arun Balaji Buduru$^{\$}$\thanks{$^{\$}$Primary Advisor}}
\address{
  IIIT-Delhi, India
}

\maketitle
 
\begin{abstract}

\noindent In this work, we introduce \textbf{\texttt{FOCA}}, a novel multimodal framework for malware classification that jointly leverages audio and visual modalities. Unlike conventional Euclidean-based fusion methods, \textbf{\texttt{FOCA}} is the first to exploit the intrinsic hierarchical relationships between audio and visual representations within hyperbolic space. To achieve this, raw binaries are transformed into both audio and visual representations, which are then processed through three key components: (i) a hyperbolic projection module that maps Euclidean embeddings into the Poincaré ball, (ii) a hyperbolic cross-attention mechanism that aligns multimodal dependencies under curvature-aware constraints, and (iii) a Möbius addition–based fusion layer. Comprehensive experiments on two benchmark datasets—Mal-Net and CICMalDroid2020--show that \textbf{\texttt{FOCA}} consistently outperforms unimodal models, surpasses most Euclidean multimodal baselines, and achieves state-of-the-art performance over existing works.

\end{abstract}



\section{Introduction}
\label{introduction}


\noindent Malware is a continuously evolving threat that compromises both system integrity and user privacy. Its classification has become a high-stakes ``cat-and-mouse'' race, with attackers deploying obfuscation and polymorphism to evade defenses. Malware family classification enhances threat understanding and strengthens incident response by linking variants to their origin and enabling targeted defenses. Researchers have explored rule-based, machine learning (ML), and deep learning (DL) approaches for malware classification, yet current methods remain vulnerable as malware continues to evolve at scale~\cite{malwarebytesreport2025}. These challenges highlight the urgent need for robust and reliable classification frameworks. The traditional approach of malware classification primarily relies on a signature-based or rule-based approach, including sample hash matching or string matching and so on, which is effective for previously seen malware, but fails when exposed to a new or obfuscated variant of it~\cite{signaturemalware}. Addressing this limitation, researchers explored static and dynamic features such as opcode n-grams, API call sequences, entropy statistics, and control flow graphs combined with ML and DL algorithms to enhance the categorization of malware samples~\cite{smith2020mlmalware, Shalaginov_2018}. In contrast, several studies have explored alternative representations of binaries by converting them into other modalities, such as \textit{binary-to-audio} or \textit{binary-to-visual (image)}, thereby enabling the application of deep learning techniques beyond raw binary analysis~\cite{kural2023apk2audio, TARWIREYI2023103282}. While image-based representations have shown competitive results in malware classification when used alone or combined with conventional features~\cite{hydra2020multimodal, narendra2023multimodal, ZHANG2025multimodal, guo2025multimodal}, audio representations remain relatively underexplored and are often restricted to spectrograms derived from traditional signal processing techniques, limiting their expressive potential. More recently, multimodal frameworks that integrate audio and visual modalities have been investigated for improved malware classification, employing fusion strategies such as cross-modal attention, feed-forward integration, or simple concatenation \cite{kaiyan2025multimodal, ismail2025midalf, LI2025multimodal}. 

However, none of the prior research has attempted to exploit the implicit hierarchical relationships between audio and visual modalities \cite{Hong_2023_ICCV} during fusion—a missed opportunity with significant potential to advance malware classification. To address this, we propose \textbf{\texttt{FOCA}} (\textbf{\texttt{F}}usi\textbf{\texttt{O}}n with Hyperbolic \textbf{\texttt{C}}ross-\textbf{\texttt{A}}ttention), a novel multimodal framework that fuses audio and visual modalities while explicitly modeling their hierarchical relationship for malware classification. For representation extraction for both modalities, we leverage state-of-the-art (SOTA) pre-trained models. Our idea is that audio modality encodes fine-grained byte-level characteristics of binary data, whereas visual modality captures higher-level spatial organization and structural patterns—together forming a hierarchy across the two modalities. To effectively model this hierarchy, \textbf{\texttt{FOCA}} makes use of hyperbolic space. Specifically, it consists of three key components: (i) projecting both modalities into a shared hyperbolic space, (ii) aligning them with a novel hyperbolic cross-attention mechanism, and (iii) merging them using Möbius addition. Extensive experiments on two benchmark datasets—Mal-Net and CICMalDroid2020—demonstrate that \textbf{\texttt{FOCA}} consistently surpasses unimodal approaches and Euclidean-based multimodal baselines, achieving SOTA performance over prior works.  To the best of our knowledge, this is the first study to leverage hyperbolic space for fusion of multimodal representations for malware classification. \noindent \textbf{To summarize, the key contributions of the proposed work are as follows}: (i) We are the first to explore the hierarchical relationship between audio and visual modalities for malware classification, (ii) We propose \textbf{\texttt{FOCA}}, a novel multimodal framework that explicitly models this hierarchy by leveraging hyperbolic space. We propose a novel hyperbolic cross-attention mechanism for aligning the audio and visual modalities. (iii) Through extensive experiments on two benchmark datasets—Mal-Net and CICMalDroid2020—\textbf{\texttt{FOCA}} consistently outperforms unimodal models and Euclidean-based multimodal baselines, achieving SOTA performance over previous works and establishing hyperbolic multimodal fusion as a promising new direction for malware classification. \textit{The code and trained models are available here: \url{https://github.com/nitinc24009/FOCA.git}.}

\section{Methodology}

In this section, we first describe the conversion of binaries into audio and visual modalities. We then present the pre-trained models (PTMs) used for extracting representations from these modalities, followed by a detailed discussion of our proposed framework, \textbf{\texttt{FOCA}}.

\subsection{Binary-to-X Transformation}

\noindent \textbf{Binary-to-Audio Modality}: Raw byte sequences from the APK-dex files are mapped into waveform samples and stored as .wav files. This representation treats binaries as continuous time-series signals, and this will enable audio PTMs to process them directly as raw waveforms.

\noindent \textbf{Binary-to-Image Modality}: The raw bytes from APK-dex files are read as unsigned integers (0–255), concatenated into a 1D array, and then padded and reshaped into a 2D matrix with a width determined by file size. The header, data, and remaining sections are mapped to the red, green, and blue channels, respectively, producing an image representation suitable for vision PTMs.

\subsection{Pre-Trained Models (PTMs)}

\noindent \textbf{Audio PTMs}: We consider SOTA audio PTMs in our study, namely, Wav2vec2~\cite{wav2vec2}, WavLM~\cite{wavlm}, and HuBERT~\cite{hubert}. Wav2vec2 was pre-trained in a self-supervised format and learns contextual representations via contrastive learning. WavLM is also trained in a self-supervised format. It solves both the masked speech prediction task and speech denoising during its pre-training. HuBERT is pre-trained in a BERT-like masked prediction objective and builds on the Wav2vec2 architecture. For Wav2vec2, WavLM, and HuBERT, we use the base versions available in Huggingface, extracting 768-dimensional representations from the frozen PTMs' last hidden layer via mean pooling.

\noindent \textbf{Vision PTMs}: We consider three SOTA vision PTMs: ResNet50~\cite{resnet-50}, VGG19~\cite{vgg-19}, and ViT~\cite{vit}. ResNet50 and VGG-19 are CNN-based PTMs pretrained on ImageNet for large-scale image classification. In contrast, ViT is a purely attention-based, convolution-free architecture. We extract 2048, 4096, and 768-dimensional representations from the last layer of frozen ResNet50, VGG19, and ViT, respectively, using average pooling.


\subsection{Unimodal Modeling}
\label{modeling}

As the downstream network for unimodal representations, we employ a CNN-based architecture. Extracted PTM representations are passed through two 1D-CNN layers (64 and 128 filters, kernel size 3), each followed by max pooling. The resulting features are flattened and fed into a fully connected layer of 128 neurons, before an output layer with softmax activation produces the class probabilities.


\begin{figure*}[hbt!]
    \centering
    \includegraphics[width=\linewidth]{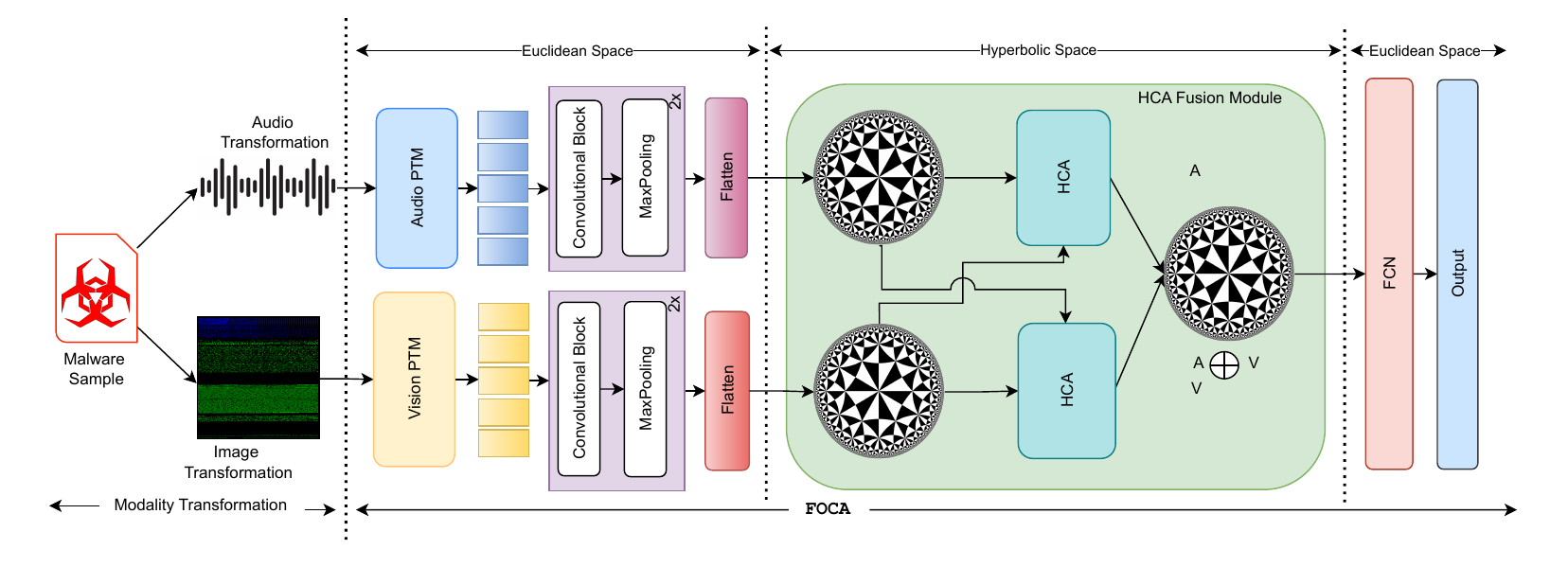}
    \caption{\texttt{\textbf{FOCA}}; $HCA$ ($\oplus$) represents Hyperbolic Cross-attention, mobius addition respectively}
    \label{fig:foca}
\end{figure*}


\subsection{\textbf{\texttt{FOCA}}}

We propose \textbf{\texttt{FOCA}}, a novel framework for the fusion of audio and visual modality representations for malware classification. The architecture is shown in Figure \ref{fig:foca}. \textbf{\texttt{FOCA}} employs a hyperbolic cross-attention (HCA) mechanism that aligns audio and visual representations while preserving the curvature constraints of hyperbolic space. HCA is introduced as a remedy for aligning modalities in hyperbolic space, whereas the vanilla cross-attention mechanism may fail to capture such dependencies. First, the extracted representations from the pre-trained models for both audio and visual inputs are passed through a CNN block with the same architectural details as described in the unimodal modeling stage. The resulting features are then flattened and projected into hyperbolic space before being processed by the HCA module. Let audio and visual features after flattening are $H^{(a)}$ and $H^{(v)}$ with $ H^{(a)}, H^{(v)} \in \mathbb{R}^{n \times d} $. These features are projected into the hyperbolic Poincare ball $\mathbb{B}^d$ using the exponential map at the origin: $ \mathcal{H}^{(a)} = \exp_{\mathbf{0}}(H^{(a)}) $ and $\mathcal{H}^{(v)} = \exp_{\mathbf{0}}(H^{(v)}) $ with
\[
\exp_{\mathbf{0}}(\mathbf{x}) = \tanh(\|\mathbf{x}\|)\frac{\mathbf{x}}{\|\mathbf{x}\|}.
\]

\noindent From these hyperbolic features, queries, keys, and values are defined as: $
\mathcal{Q}^{(a)}, \mathcal{K}^{(a)}, \mathcal{V}^{(a)} \in \mathbb{B}^d $ and $ \mathcal{Q}^{(v)}, \mathcal{K}^{(v)}, \mathcal{V}^{(v)} \in \mathbb{B}^d $. Following this, we calculate the HCA. Cross-modal attention weights are computed using hyperbolic distance:
\[
\alpha^{a \rightarrow v}_{ij} = 
\frac{\exp\!\big(-d_{\mathbb{H}}(\mathcal{Q}^{(a)}_i, \mathcal{K}^{(v)}_j)\big)}
{\sum_{j'} \exp\!\big(-d_{\mathbb{H}}(\mathcal{Q}^{(a)}_i, \mathcal{K}^{(v)}_{j'})\big)},
\]
\[
\alpha^{v \rightarrow a}_{ij} = 
\frac{\exp\!\big(-d_{\mathbb{H}}(\mathcal{Q}^{(v)}_i, \mathcal{K}^{(a)}_j)\big)}
{\sum_{j'} \exp\!\big(-d_{\mathbb{H}}(\mathcal{Q}^{(v)}_i, \mathcal{K}^{(a)}_{j'})\big)},
\]
\noindent where the hyperbolic distance in the Poincar\'e ball is
\[
d_{\mathbb{H}}(\mathbf{x},\mathbf{y}) = 
\operatorname{arcosh}\!\left(1 + 
\frac{2\|\mathbf{x}-\mathbf{y}\|^2}{(1-\|\mathbf{x}\|^2)(1-\|\mathbf{y}\|^2)}\right).
\]

\noindent After this, the attended values are aggregated via möbius operations:
\[
\mathcal{O}^{a \rightarrow v}_i = \bigoplus_{j}\, \alpha^{a \rightarrow v}_{ij} \otimes_{\mathbb{H}} \mathcal{V}^{(v)}_j,
\quad
\mathcal{O}^{v \rightarrow a}_i = \bigoplus_{j}\, \alpha^{v \rightarrow a}_{ij} \otimes_{\mathbb{H}} \mathcal{V}^{(a)}_j,
\]
where mobius addition $\mathbf{x} \oplus_{\mathbb{H}} \mathbf{y}$ and mobius scalar multiplication $r \otimes_{\mathbb{H}} \mathbf{x}$ is given as:

\[
\mathbf{x} \oplus_{\mathbb{H}} \mathbf{y} =
\frac{(1+2\langle \mathbf{x},\mathbf{y}\rangle + \|\mathbf{y}\|^2)\mathbf{x} + (1-\|\mathbf{x}\|^2)\mathbf{y}}
{1+2\langle \mathbf{x},\mathbf{y}\rangle + \|\mathbf{x}\|^2\|\mathbf{y}\|^2},
\]

\[
r \otimes_{\mathbb{H}} \mathbf{x} =
\tanh\!\big(r \tanh^{-1}(\|\mathbf{x}\|)\big) \frac{\mathbf{x}}{\|\mathbf{x}\|}.
\]
\noindent The outputs from both directions are fused in hyperbolic space using mobius addition and then mapped back to euclidean space:$
\mathbf{O} = \log_{\mathbf{0}}\!\Big(\mathcal{O}^{a \rightarrow v} \oplus_{\mathbb{H}} \mathcal{O}^{v \rightarrow a}\Big) $ with the logarithmic map at the origin:
\[
\log_{\mathbf{0}}(\mathbf{x}) = \tanh^{-1}(\|\mathbf{x}\|)\frac{\mathbf{x}}{\|\mathbf{x}\|}.
\]
 
\noindent The fused representation $\mathbf{O}$ is then passed through fully connected layers with 120 and 30 neurons followed by output layer with softmax activation for malware classification. The number of trainable parameters ranges from 2.7M to 4.5M depending on the input representations dimensions.

\section{Experiments}
\label{experiments}


\subsection{Dataset}

\noindent \textbf{CICMalDroid-2020}~\cite{mahdavifar2022effective, mahdavifar2020dynamic}: It is an open-access dataset that is a collection of 17341 APK samples that were collected and curated by the Canadian Institute of Cybersecurity, collected between December 2017 and December 2018. This dataset comprises one benign category and four distinct categories of malware: Adware, Banking malware, SMS malware, and Riskware.

\noindent \textbf{Mal-Net~\cite{freitas2022malnetlargescaleimagedatabase}}: It is a public dataset that contains more than 1.2 million images that comprise 47 types and 696 families of malware. We filtered out a total of eight thousand samples from ten different classes: adware, clicker+trojan, spyware, addisplay, smssend, monitor, ransom+trojan, banker+trojan, adware+trojan, and rogue. We selected these categories based on the sample size of each category, and selected 800 samples randomly from each category to balance the dataset. The respective APK files have been sourced by mapping the Mal-Net signatures with the Androzoo~\cite{androzoo} dataset. 

\noindent \textbf{Training Details:} We consider categorical cross-entropy as the loss function, Adam as optimizer, batch size to be 32, and set the learning rate to be 1e-5 for 50 epochs for training the model. We incorporate dropout and early stopping to reduce the chances of overfitting. We follow a 5-fold cross-validation strategy for training and validating the performance by considering 4 folds for training and 1 fold for testing.

\begin{table}[ht!]
    \scriptsize
    \centering
    \caption{Performance of unimodal and multimodal models; 
    Accuracy and macro-F1 are reported in \%; Here, $+$ and $\otimes$ represent concatenation and cross-modal attention in Euclidean space, and $\boxplus$ defines hyperbolic cross-attention fusion via \textbf{\texttt{FOCA}}, respectively; The scores are average of five folds; The results in \textcolor{blue}{blue} color define reproducing the algorithms and testing on our end}
    \label{tab:results}
        \begin{tabular}{l|cc|cc}
        \toprule
        \multirow{2}{*}{Model} &
        \multicolumn{2}{c|}{Mal-Net} &
        \multicolumn{2}{c}{CICMalDroid2020} \\
        \cmidrule(lr){2-3} \cmidrule(lr){4-5}
        & Accuracy & macro-F1 & Accuracy & macro-F1 \\
        \midrule
        \multicolumn{5}{c}{\textbf{Audio modality}} \\
        \midrule
        WavLM   & 63.31 & 60.22 & 73.69 & 71.71 \\
        Wav2Vec2  & \cellcolor{darkgreen}68.82 & \cellcolor{darkgreen}65.46 & \cellcolor{lightgreen}76.12 & \cellcolor{lightgreen}74.07 \\
        HuBERT   & \cellcolor{lightgreen}64.69 & \cellcolor{lightgreen}61.53 & \cellcolor{darkgreen}80.98 & \cellcolor{darkgreen}78.80 \\
        \midrule
        \multicolumn{5}{c}{\textbf{Image modality}} \\
        \midrule
        ViT & \cellcolor{darkgreen}62.90 & \cellcolor{darkgreen}58.48 & \cellcolor{darkgreen}74.90 & \cellcolor{darkgreen}74.48 \\
        VGG-19 & \cellcolor{lightgreen}61.01 & \cellcolor{lightgreen}56.73 & 72.65 & 72.25 \\
        ResNet-50 & 59.87 & 55.61 & 71.18 & 70.76 \\
        \midrule
        \multicolumn{5}{c}{\textbf{Multi-modality (Audio $+$ Image)}} \\
        \midrule
         WavLM  $+$ ViT   & \cellcolor{lightgreen}71.20 & \cellcolor{lightgreen}67.61 & 79.74 & 79.44 \\
         WavLM  $+$ VGG-19   & 69.73 & 66.22 & 78.53 & 78.23 \\
         WavLM  $+$ ResNet-50   & 70.46 & 66.91 & 78.10 & 77.83 \\
         Wav2Vec2 $+$ ViT   & \cellcolor{lightgreen}71.93 & \cellcolor{lightgreen}68.31 & \cellcolor{darkgreen}82.21 & \cellcolor{darkgreen}81.90 \\
         Wav2Vec2 $+$ VGG-19   & 70.98 & 67.43 & 78.92 & 78.63 \\
         Wav2Vec2 $+$ ResNet-50   & \cellcolor{darkgreen}73.40 & \cellcolor{darkgreen}69.70 & 79.74 & 79.45 \\
         HuBERT  $+$ ViT   & 70.66 & 67.03 & \cellcolor{lightgreen}80.16 & \cellcolor{lightgreen}79.85 \\
         HuBERT  $+$ VGG-19   & 70.03 & 66.41 & 78.92 & 78.65 \\
         HuBERT  $+$ ResNet-50   & 70.19 & 66.53 & 78.10 & 77.85 \\
        \midrule
        \multicolumn{5}{c}{\textbf{Multi-modality (Audio $\otimes$ Image)}} \\
        \midrule
         WavLM  $\otimes$ ViT   & \cellcolor{lightgreen}74.11 & \cellcolor{lightgreen}72.22 & \cellcolor{lightgreen}85.56 & \cellcolor{lightgreen}85.25 \\
         WavLM  $\otimes$ VGG-19   & 72.94 & 71.03 & 84.23 & 83.94 \\
         WavLM  $\otimes$ ResNet-50   & 73.29 & 71.35 & 83.81 & 83.51 \\
         Wav2Vec2 $\otimes$ ViT   & \cellcolor{lightgreen}74.86 & \cellcolor{lightgreen}72.89 & \cellcolor{darkgreen}93.21 & \cellcolor{darkgreen}91.90 \\
         Wav2Vec2 $\otimes$ VGG-19   & 73.69 & 71.78 & 88.68 & 84.37 \\
         Wav2Vec2 $\otimes$ ResNet-50   & 74.48 & 72.53 & 87.52 & 85.21 \\
         HuBERT  $\otimes$ ViT   & \cellcolor{darkgreen}76.78 & \cellcolor{darkgreen}74.77 & \cellcolor{lightgreen}92.21 & \cellcolor{lightgreen}91.89 \\
         HuBERT  $\otimes$ VGG-19   & 74.01 & 72.13 & 84.75 & 84.42 \\
         HuBERT  $\otimes$ ResNet-50   & 73.71 & 71.78 & 83.94 & 83.63 \\
         \midrule
         \multicolumn{5}{c}{\textbf{Multi-modality (Audio $\boxplus$ Image)}} \\
         \midrule
         WavLM  $\boxplus$ ViT   & \cellcolor{lightgreen}79.73 & \cellcolor{lightgreen}80.10 & 88.47 & 87.23 \\
         WavLM  $\boxplus$ VGG-19   & 78.84 & 78.18 & 86.65 & 85.41 \\
         WavLM  $\boxplus$ ResNet-50   & 78.35 & 78.90 & 87.56 & 86.34 \\
         Wav2Vec2 $\boxplus$ ViT   & \cellcolor{medgreen}80.98 & 77.98 & \cellcolor{lightgreen}91.21 & \cellcolor{lightgreen}89.90 \\
         Wav2Vec2 $\boxplus$ VGG-19   & 79.27 & 78.70 & 87.68 & 86.43 \\
         Wav2Vec2 $\boxplus$ ResNet-50   & 79.56 & 79.56 & 88.62 & 87.35 \\
         HuBERT $\boxplus$ ViT   & \cellcolor{darkgreen}\textbf{82.84} & \cellcolor{darkgreen}\textbf{81.72} & \cellcolor{darkgreen}\textbf{99.10} & \cellcolor{darkgreen}\textbf{98.85} \\
         HuBERT $\boxplus$ VGG-19   & 80.16 & 80.40 & 86.65 & 85.42 \\
         HuBERT $\boxplus$ ResNet-50   & 79.27 & 80.20 & 87.56 & 86.31 \\
        \midrule
        \multicolumn{5}{c}{\textbf{Comparison to SOTA}} \\
        \midrule
        Samaneh et al.~\cite{mahdavifar2020dynamic} & \textcolor{blue}{41.73} & \textcolor{blue}{40.84} & 96.73 & 97.84 \\
        Scott et al.~\cite{freitas2022malnetlargescaleimagedatabase} & \cellcolor{darkgreen}70.12 & \cellcolor{darkgreen}67.71 & \textcolor{blue}{93.74} & \textcolor{blue}{91.81} \\
        Devnath et al.~\cite{devnath2025icccn} & 52.49 & 51.21 & \textcolor{blue}{92.74} & \textcolor{blue}{0.9383} \\
        Vasan et al.~\cite{vasan2025advanced} & \textcolor{blue}{48.54} & \textcolor{blue}{51.67} & 95.71 & 96.46 \\
        Yang et al.~\cite{yang2024pvitnet} & \cellcolor{darkgreen}\textcolor{blue}{53.71} & \cellcolor{darkgreen}\textcolor{blue}{53.65} & \cellcolor{darkgreen}98.52 & \cellcolor{darkgreen}98.31 \\
        \bottomrule
    \end{tabular}
\end{table}

\subsection{Results and Discussion}

\noindent Table~\ref{tab:results} reports the unimodal results for the audio and image modalities. Overall, audio-based representations outperform image-based representations across both datasets. Within the audio modality, no single representation emerges as a consistent winner, as Wav2vec2 performs better on one dataset while HuBERT achieves stronger results on the other. For the image modality, ViT provides the best performance among the tested visual encoders. Table~\ref{tab:results} compares different fusion strategies: $x + y$ denotes simple concatenation, $x \otimes y$ represents cross-attention in euclidean space, and $x \boxplus y$ corresponds to \textbf{\texttt{FOCA}}. Here, $x$ and $y$ denote any pair of audio and visual representations. We additionally consider simple concatenation and Euclidean cross-attention as baselines. For simple concatenation, modality features are concatenated after the flatten layer, following the same architecture as \textbf{\texttt{FOCA}}. For euclidean cross-attention, we adopt the same structure but replace hyperbolic cross-attention with vanilla euclidean attention. All models are trained under identical settings to ensure fair comparison. We observe that Euclidean cross-attention consistently outperforms simple concatenation, owing to its stronger alignment capability. This trend is evident across most audio–visual representation pairs; for instance, HuBERT $+$ ViT achieves 70.66\% accuracy, whereas HuBERT $\otimes$ ViT improves performance to 76.78\%. Overall, the fusion of audio and visual modalities through \textbf{\texttt{FOCA}} yields the best performance compared to unimodal approaches as well as baseline fusion techniques. This highlights the effectiveness of \textbf{\texttt{FOCA}} in integrating audio and visual representations while explicitly modeling the hierarchical relationships between them in hyperbolic space. The consistent improvements across diverse representation pairs further demonstrate that curvature-aware alignment offers a principled advantage over euclidean fusion. We got the best performance with HuBERT and ViT through \textbf{\texttt{FOCA}}. Additionally, the variances in the performance observed among different representations is mostly due to the downstream data distribution effecting the performance. We also present t-SNE visualizations in Figure~\ref{fig:tsne} for euclidean cross-attention (HuBERT $\otimes$ ViT) and \textbf{\texttt{FOCA}} (HuBERT $\boxplus$ ViT), where the representations are extracted from the penultimate layer. Compared to Euclidean cross-attention, FOCA produces more distinct and compact clusters, illustrating its effectiveness. Furthermore, as shown in Table~\ref{tab:results} (Comparison to SOTA), \textbf{\texttt{FOCA}} establishes new SOTA on both datasets surpassing previous SOTA works.

\begin{figure}[ht]
    \centering
    \begin{subfigure}{0.23\textwidth}
        \centering
        \includegraphics[width=\linewidth]{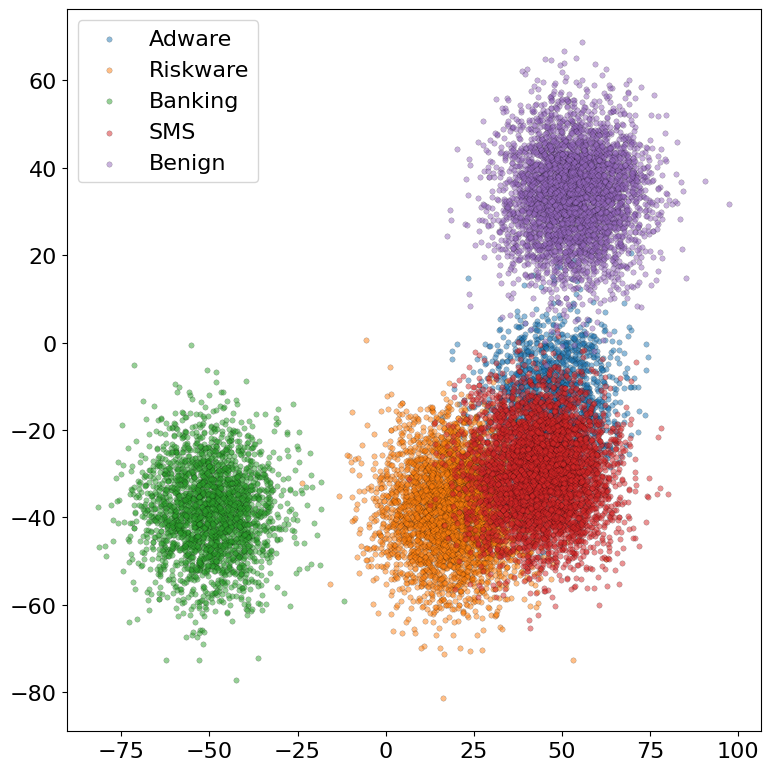}
        \caption{}
        \label{fig:tsne_hubert}
    \end{subfigure}
    \begin{subfigure}{0.23\textwidth}
        \centering
        \includegraphics[width=\linewidth]{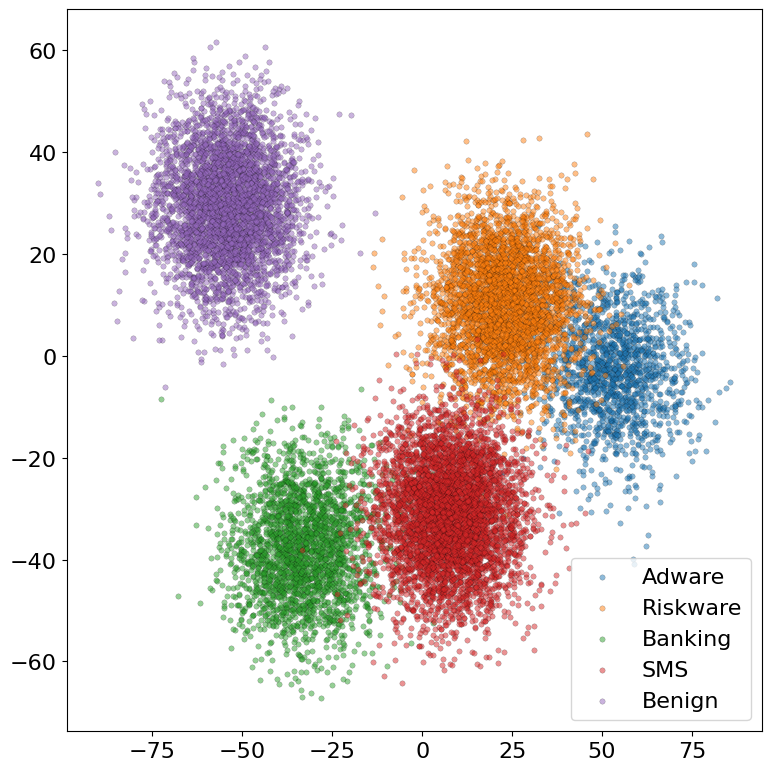}
        \caption{}
        \label{fig:tsne}
    \end{subfigure}
    \caption{\textit{t}-SNE visualizations: (a) Euclidean Cross-Attention (HuBERT  $\otimes$ ViT) (b) \textbf{\texttt{FOCA}} (HuBERT $\boxplus$ ViT) on MalDroid-2020 Dataset}
    \label{fig:tsne}
\end{figure}



\vspace{-1em}

\section{Conclusion}
\label{conclusion}

\vspace{-1em}

We introduced \texttt{\textbf{FOCA}}, a novel multimodal malware classification framework that leverages audio representations and visual representations of binaries and fuses them in hyperbolic space through hyperbolic cross-attention. Empirical evaluation of \texttt{\textbf{FOCA}} on two benchmark datasets, CICMalDroid-2020 and MalNet, shows that our method achieves the top performance across competitive baseline and sets new SOTA over previous works. By bridging hierarchical relationships modeling with geometry-aware approaches, this work opens a new direction for designing effective malware classification systems and a sets the path for future research.

\bibliographystyle{IEEEbib}
\bibliography{main}

\end{document}